\documentclass[preprint, APS,showpacs, preprintnumbers,amsmath]{revtex4}
\usepackage{graphicx}
\usepackage{dcolumn}
\usepackage{bm}
\usepackage{wrapfig}
\usepackage{amssymb}
\usepackage{amsmath}
\def\be{\begin{equation}}
\def\ee{\end{equation}}
\def\e#1{\label{#1}\end{equation}}
\def\bea{\begin{eqnarray}}
\def\eea{\end{eqnarray}}
\def\ea#1{\label{#1}\end{eqnarray}}

\def\bem#1{\begin{mathletters}\label{#1}}
\def\eml{\end{mathletters}}

\def\4#1{{\boldsymbol{#1}}}
\def\8#1{{\widetilde{#1}}}
\def\bse{\begin{subequations}}
\def\ese{\end{subequations}}
\def\nn{\nonumber}
\begin{document}
\title{Effect of qubit losses on Grover's quantum search algorithm}
\author{D. D. Bhaktavatsala Rao}
\author{Klaus M{\o}lmer}
\affiliation{
Lundbeck Foundation Theoretical Center for Quantum System Research, Department of Physics and Astronomy,
University of Aarhus, DK-8000 Aarhus C, Denmark}
\date{\today}

\begin{abstract}
We investigate the performance of Grover's quantum search algorithm on a register which is subject to loss of the particles that carry the qubit information. Under the assumption that the basic steps of the algorithm are applied correctly on the correspondingly shrinking register, we show that the algorithm converges to mixed states with 50\% overlap with the target state in the bit positions still present. As an alternative to error correction, we present a procedure that combines the outcome of different trials of the algorithm to determine the solution to the full search problem. The procedure may be relevant for experiments where the algorithm is adapted as the loss of particles is registered, and for experiments with Rydberg blockade interactions among neutral atoms, where monitoring of the atom losses is not even necessary.
\end{abstract}

\maketitle

\section{Introduction}
The quantum search protocol over an unstructured database of $N = 2^n$ elements ($n$ is the number of qubits) identifies the correct $n$-bit string that matches a given target state. The search algorithm originally proposed by Grover \cite{grov} is an iterative algorithm, where each step in the algorithm involves two operations: A change of sign of the target state amplitude and an inversion of all quantum state amplitudes about their mean in the computational basis \cite{grov,gengrov}. The ideal evolution during a single iteration of the algorithm can be described by a rotation in the two-dimensional subspace spanned by the target state and the symmetric superposition of all logical states. This explains the functioning of the algorithm and allows exact evaluation of the unitary dynamics. After $\frac{\pi}{4} \sqrt{N}$ iteration steps, the algorithm has rotated an initially symmetric superposition of all computational basis states of the qubit register into a state very close to the target state. The advantage of Grover's algorithm (GA) over a classical search algorithm is obtained only when the size of the search space is large ($N \gg 1$). The larger the database, the longer, however, is the time ($\sim \sqrt{N}$) for finding the target state, and the more important are the effects of error processes relevant to the physical system \cite{exp1,exp2,exp3,exp4,exp5}.
While error processes that preserve the two-level picture of the quantum search have been studied by many authors \cite{noise1, noise2}, single qubit dephasing and dissipation leak population out of this subspace and one has to resort to numerical methods for estimating the success probability of the algorithm \cite{num1,num2,num3}.

An important error process in some physical schemes for quantum computing is the physical loss of the particles that carry the qubit information. Such errors are relevant for neutral atom-based quantum computing protocols, where losses due to collisions with the background gas and due to the heating by laser excitation are common at long time scales \cite{laser1,laser2}. Schemes to identify and correct errors due to losses have received some attention. The fact that we may know precisely which qubit is lost and has to be replaced leads to simpler schemes and quite high error thresholds for fault tolerant quantum computing compared to the case of general qubit errors \cite{varnava,ralph,stace,herrera}.  In this work we shall provide a full theoretical analysis of Grover's algorithm in the presence of random qubit losses during the search operation. In our analysis we shall make no attempts to detect or correct particle losses. Instead, we will assume an implementation of the algorithm where the sign changes are implemented perfectly on the relevant Hilbert space of the remaining qubits, i.e., a sign change is made on the state component with correct bit values in all the positions of the remaining qubits and the amplitude inversion about the mean refers to the state amplitudes in the computational tensor product basis of the remaining qubits. As we will show, the Grover steps proceed while the system passes through a sequence of two-dimensional subspaces of the shrinking tensor product spaces, and eventually we end up with a candidate solution for a subset of the target bit values. We suggest an effective method by which one can reconstruct the full target state after a modest number of independent trials have provided candidate solutions for all bit values.

The paper is structured as follows. In Sec. II, we shall describe the effect of qubit loss as a transition between two-dimensional subspaces of the tensor product spaces of different numbers of qubits. In Sec. III, we describe the Grover steps by an effective Hamiltonian. We then derive a continuous time master equation which describes the combined effect of the algorithm and the random loss process and we analyze the resulting dynamics. In Sec. IV, we describe a method to efficiently obtain the solution of the search problem from different subsets of surviving qubits in independent trials of the algorithm. In Sec. V, we describe a particular physical implementation of quantum computing using neutral atoms and Rydberg blockade interactions, for which the present analysis applies. Sec. VI concludes the article.

\begin{figure}
\includegraphics[height=60mm,width=120mm]{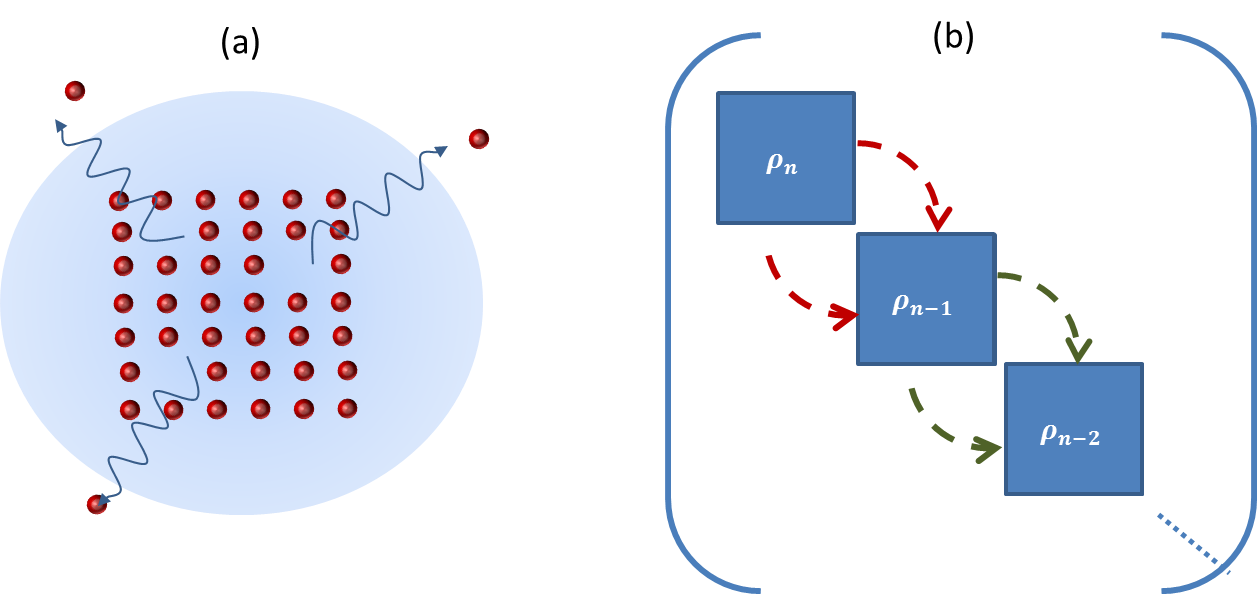}
\caption{(a). Illustration of physical loss of qubits from a quantum register.
(b). Schematic representation of the block-diagonal density matrix, with population leaking from one two-dimensional subspace to the other caused by the loss of qubits.}
\end{figure}

\section{Grover algorithm in the presence of losses}

In this section, we show explicitly that the Grover algorithm, both in the absence and in the presence of losses can be described by quantum dynamics within two-dimensional subspaces. Each loss event merely shifts the dynamics from one $2D$ space to another. Henceforth, the total evolution will be described by a concatenated evolution among these subspaces. Since the remaining qubits are entangled with the lost ones, they are properly described by a mixed state with a corresponding reduced density matrix.
We thus loose both the information carried by the lost qubits and the purity of the state of the remaining qubits in any given run (trial) of the algorithm.

The quantum data-base search algorithm initially operates in an $N=2^n$-dimensional Hilbert space, where the basis vectors are represented as $n$-bit binary strings, $|x_n\rangle = |b_{x,1},b_{x,2}\cdots b_{x,n}\rangle$, where every $b_{x,m}$ takes two values $0$ and $1$. The aim of the Grover search algorithm is to map an equal superposition state of all $N$ basis vectors into a definite target object $|\tilde{x}_n\rangle = |b_{1},b_{2}\cdots b_{n}\rangle$.

The search process on the $n$-qubit system is carried out by repeated action of a unitary operator $G_n$ which performs a change of sign of the target state amplitude and an inversion of all basis state amplitudes about their mean value \cite{grov}.
The symmetry of the Grover operator $G$ implies that the dynamics is restricted to a $2D$ Hilbert space spanned by the target state $|\tilde{x}_n\rangle$ and the fully symmetric state $\frac{1}{\sqrt{N}}\sum_{x_n} |x_n\rangle$. As the target state is one of the terms in the sum over all basis states, it is convenient to span the 2D space by the orthogonal basis formed by the target state $|\tilde{x}_n\rangle$ and $|{s}_n\rangle \equiv \frac{1}{\sqrt{N-1}}\sum_{x_n\ne \tilde{x}_n}|x_n\rangle$.
In this basis, $G_n$ amounts to a rotation by an angle $\theta_n= 2\sin^{-1}(1/\sqrt{N})$,
\be
\label{gop}
G_n = \cos{\theta_n}I - i\sin{\theta_n}\sigma_y,
\ee
where $\sigma_y$ and $I$ are the Pauli-$y$ matrix and the identity operator respectively. If the initial state $\rho_n(0)$ of the $n$-qubit register can be described within the $2D$ subspace, then $k$ applications of $G_n$ yields the state
\be
\label{rhk}
\rho_{n}(k) = G^k_n\rho_{n}(0)(G^k_n)^\dagger.
\ee
A complete transition from the symmetric state (which is readily prepared as a product state with every register qubit in the even superpositions state $(|0\rangle +|1\rangle)/\sqrt{2}$)  to the target state is obtained by
$\frac{\pi}{4}\sqrt{N}$ consecutive rotations, i.e., repeated operations of $G_n$. This leads to the square root speed-up compared to a classical search through all $N$ elements.

\subsection{Losses}
We now consider the situation where a number of qubits are lost after $k$ applications of the Grover operator. If $m$ qubits remain after the losses, the subsequent evolution takes place within a Hilbert space of dimension $M=2^m$. Using the fact that both the target state and the fully symmetric state factorize, $|\tilde{x}_n\rangle = |\tilde{x}_{m}\rangle\otimes|\tilde{x}_{n-m}\rangle$,  $\frac{1}{\sqrt{N}}\sum_{x_n} |x_n\rangle= \frac{1}{\sqrt{M}}\sum_{x_{m}}|x_{m}\rangle\otimes \frac{1}{\sqrt{N/M}}\sum_{x_{n-m}}|x_{n-m}\rangle $, we can  rewrite the basis vector $|s_n\rangle =\frac{1}{\sqrt{N-1}}\left [\sum_{x_{n}}|x_{n}\rangle-|\tilde{x}_{n}\rangle \right] $ in terms of the corresponding basis vectors of the subspaces containing $m$ and $n-m$ qubits,
\bea
\label{basis}
|s_n\rangle =f_{n,m}|s_{m}\rangle\otimes|s_{n-m}\rangle + g_{n,m}|\tilde{x}_{m}\rangle\otimes|s_{n-m}\rangle + h_{n,m}|s_{m}\rangle\otimes|\tilde{x}_{n-m}\rangle,
\eea
where $f_{n,m} = \sqrt{\frac{(N-M)(M-1)}{M(N-1)}}$, $g_{n,m} = \sqrt{\frac{N-M}{M(N-1)}}$ and $h_{n,m}=\sqrt{\frac{M-1}{N-1}}$. It thus follows, that after discarding $n-m$ qubits, the reduced density matrix of the remaining $m$ qubits is given by
\be
\label{ptrh}
\rho_m(k) \equiv Tr_{n-m}[\rho_n(k)] = \langle \tilde{x}_{n-m}|\rho_n(k)|\tilde{x}_{n-m} \rangle + \langle s_{n-m}|\rho_n(k)|s_{n-m} \rangle.
\ee
The key observation is that even though the system loses qubits, its effective dynamics can still be described by a density matrix evolution in the two-dimensional space spanned by the basis vectors $|\tilde{x}_{m}\rangle$ and $|s_{m}\rangle$.

After performing the partial trace one finds that the elements $\rho^{11}_m(k) = \langle \tilde{x}_m|\rho_m(k)|\tilde{x}_m \rangle$,
$\rho^{12}_m(k) = \langle s_m|\rho_m(k)|\tilde{x}_m \rangle$ and $\rho^{22}_m(k) = \langle s_m|\rho_m(k)|s_m \rangle$ of the reduced density matrix $\rho_m(k)$ can be expressed in terms of the elements of the density matrix $\rho_n(k)$ as follows:
\bea
\label{rhcomp}
\rho^{11}_m(k) &=& \rho^{11}_n(k) + g^2_{n,m}\rho^{22}_n(k), ~
\rho^{22}_m(k) = (1-g^2_{n,m})\rho^{22}_n(k), \\ \nn
\rho^{12}_m(k) &=& {h_{n,m}}\rho^{12}_n(k) + f_{n,m}g_{n,m} \rho^{22}_n(k).
\eea
Since the lost qubits are entangled with the register one always finds, $Tr[\rho^2_m(k)] <Tr[\rho^2_n(k)]$, i.e., the reduced density matrix of the surviving qubits is always mixed.

\section{Master equation}

As the system loses the qubits one by one in a random manner, the actual number of remaining qubits at any given time becomes a random variable, and our knowledge about the system is appropriately described by a weighted sum over corresponding two-dimensional density matrices. We take the duration of a full Grover step as unit of time, and since each Grover step implements a small angular rotation (Eq. (\ref{gop})), we treat this as an effective continuous rotation in each $m$-qubit subspace, driven by the Hamiltonian $H_m=\omega_m \sigma_y$, where $\omega_m = 2/\sqrt{M}=2^{-(m-2)/2}$ accounts for the rotation angle per time step in the space of $m$-qubits, and we set $\hbar=1$. The losses constitute a continuous random process with an average feeding of probability towards states with smaller qubit numbers.

Let $\gamma$ denote the (small) probability for any qubit to be lost during the duration of a single Grover step. Due to the small values of the Grover rotation angle and the loss probability, we will represent both processes by a continuous time master equation
\be
\label{msteq}
\frac{\partial\rho_{m}(t)}{\partial t}= -i[H_m,\rho_m(t)] - m\gamma \rho_m(t) + (m+1)\gamma\mathcal{L}(\rho_{m+1}(t))
\ee
where $\rho_{m}(t)$ represents the projection of the density matrix on the subspace with $m$ surviving qubits, and $(m+1)\gamma\mathcal{L}(\rho_{m+1}(t))$ represents the map in
Eq. (\ref{rhcomp}) for the loss of a single qubit, $n=m+1$. The coupling due to losses of different projections of the density matrix is illustrated in Fig 1b.

The effect of the loss operator coupling the $m^{th}$ and the $m+1^{st}$ subspace density matrices can be found explicitly using the above equations,
\bea
\label{mastcomp}
\frac{\partial p_{m}(t)}{\partial t} & = &  -m\gamma p_{m}(t) + (m+1)\gamma p_{m+1}(t) \\
\frac{\partial \rho^{w}_m(t)}{\partial t}  &=& 2\omega_m \rho^{u}_m -m\gamma\rho^{w}_m(t) + (m+1)\gamma ((1-A_{m})\rho^{w}_{m+1}(t))+A_{m} p_{m+1}(t))\nn \\
\frac{\partial \rho^{u}_m(t)}{\partial t} &=& -2\omega_m \rho^{w}_m - m\gamma\rho^{u}_m(t) + (m+1)\gamma (B_{m}\rho^{u}_{m+1}(t) -\frac{B_{m}}{2\sqrt{2M-1}}[\rho^{w}_{m+1}(t)-p_{m+1}(t)])  \nn
\eea
where, $p_m(t) = Tr[\rho_{m}(t)]$, $\rho^{w}_m(t) = \rho^{11}_m(t)-\rho^{22}_m(t)$ gives the population difference between the target and the symmetric states $|\tilde{x}_m\rangle$, $|s_m \rangle$, and $\rho^{u}_m(t) = (\rho^{12}_m(t)+\rho^{21}_m(t))/2$, gives the real part of their coherence in each subspace.
The arguments leading to Eq. (\ref{rhcomp}) yield, for the loss of a single qubit, the coefficients $A_m = \frac{1}{2M+1}$ and $B_m = \sqrt{\frac{M-1}{2M-1}}$ in
Eq. (\ref{mastcomp}).

Note that we used the same symbol for the density matrix in
Eqs. (\ref{ptrh}, \ref{rhcomp}) which was derived for a definite number of qubits and therefore has unit trace, while the first equation in Eq. (\ref{mastcomp}) for the weight distribution $p_m(t)$ leads, irrespective of the Grover dynamics, to the binomial distribution for the surviving number of qubits,
\label{wtdst}
\be
p_m(t) = \binom{n}{m}
 e^{-m\gamma t}(1-e^{-\gamma t})^{n-m}.
\ee
We now address the question of how much population has been transferred to the target state within the different subspaces, since this will tell us with what probability one obtains correct outcomes upon readout of the qubits remaining at the end of the calculation.

Since there are only $2n$ coupled equations, one can solve the dynamics exactly by numerical integration for even very large numbers of qubits. The parameter regime of interest is one, where we may loose a significant fraction of the qubits, but of course not all of them, during execution of the Grover algorithm. Such a case is illustrated in Fig. 2, where we start with 24 qubits, and a loss rate leaving only 7-8 qubits at the anticipated end of the Grover search after $\sim 3217$ Grover operations. The figure shows that after about $\sim 1300$ Grover steps, half of the qubits are still present, and the different two-dimensional subspaces are approximately populating the fully mixed states, with equal weight on the target state and on the uniform state with the corresponding number of qubits,
\be
\label{fsol}
\rho(t) \approx\sum_{m=1}^{n}p_m(t)\frac{1}{2}\left[|\tilde{x}_m\rangle\langle \tilde{x}_m| +|s_m\rangle\langle s_m|\right].
\ee
Reading out the register at this time, thus provides on average $\sim n/2$ bit values, which are all correct with a nearly $50\%$ probability (and which attain random and uncorrelated values with the remaining probability).

Note that the number of surviving qubits is governed by the decay rate $\gamma$, while the approach towards the even mixture of the target and symmetric states of the surviving qubits is governed, in parts, by the $m$-dependent $\omega_m$ and the relaxation terms in Eq. (\ref{mastcomp}) due to the trace over the lost qubits.

To estimate this latter dynamics, we note that the binomial distribution is narrow with respect to the $m$-dependence of the coefficients in  Eq. (\ref{mastcomp}), and hence, we may evaluate the average dynamics of the $m$-qubit states by considering
\be
\rho^{w}(t)=\sum_m \rho_m^{w}(t),\ \  \rho^{u}(t)=\sum_m \rho_m^{u}(t),
\ee
and evaluating the different $m$-dependent coefficients in Eq. (\ref{mastcomp}) at the time dependent average value
\be
\bar{m}(t) \equiv \sum_m mp_m(t)  = n e^{-\gamma t}.
\ee
Summing over all $m$-subspaces and assuming $M \gg 1$, such that one can approximate the coefficients, $A_m \approx 0$ and $B_m \approx 1/\sqrt{2}$ in Eq. (\ref{mastcomp}), one obtains the average dynamics, governed by the two coupled equations
\bea
\frac{\partial \rho^{w}(t)}{\partial t}  &=& 2\omega(t) \rho^{u} \nn \\
\frac{\partial \rho^{u}(t)}{\partial t}  &=& -2\omega(t)\rho^{w} -\Gamma(t)\rho^{u}.
\eea
These are the coupled equations for a two-level system with a time dependent Rabi-flopping rate,
$\omega(t) = 2^{-(\bar{m}(t)-2)/2}$ and dephasing rate $\Gamma(t) = [(\bar{m}(t)+1)B_{\bar{m}(t)} -{\bar{m}(t)}]\gamma \simeq \frac{\gamma}{\sqrt{2}} + \left(1-\frac{1}{\sqrt{2}}\right)n\gamma e^{-\gamma t}$.

Clearly if the Rabi-flopping is allowed to act long enough to distribute the population evenly among the states, before all the qubits have been lost, the state of the remaining qubits is the fully mixed state, with a $50 \%$ probability that all the bit values will be read at the correct target values.

\begin{figure}
\includegraphics[height=70mm,width=120mm]{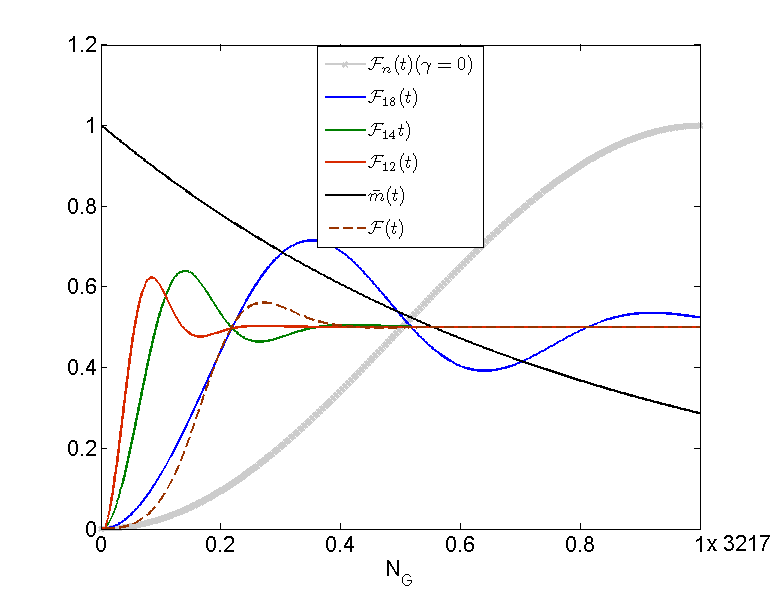}
\caption{The relative probability $\mathcal{F}_m(t) = \langle \tilde{x}_m|\rho_m|\tilde{x}_m\rangle/p_m(t)$ of the target state component in subspaces with 12 (red, steepest rise), 14 (green, second steepest rise) and 19 (blue, third steepest rise at small $N_G$) remaining qubits (out of $n=24$), plotted as functions of the number of Grover steps $N_G$. The brown, dashed line shows the weighted probability $\mathcal{F}(t) = \sum_m p_m\mathcal{F}_m(t)$ that all bits match the target bit values. The results are shown for a loss rate of $\gamma \approx 4\times 10^{-4}$ (per Grover step) leading to the average fraction of surviving qubits  shown with the black line, while the target state probability in the ideal lossless case is shown with the gray line, reaching unity for $N_G\simeq 3217$.}
\end{figure}

\section{Reconstruction of the target state}

From the above analysis it can be seen that we may loose a significant fraction of the qubits during the Grover search, and still, we may pick a moment in time, where, e.g., half of the qubits are left, and where, there is a $50\%$ chance that their states all match the corresponding bit values in the correct target state. The probability that any single qubit is measured in its correct state is therefore $75\%$, and by repeating the protocol a number of times, so that we accumulate multiple read out values for every bit, a simple majority vote will yield the target bit values.

We can, however, improve the statistical certainty of the result by exploiting the fact that with $50 \%$ probability, all the measured bit values are correct, while with the remaining $50 \%$ probability they are all random. This implies that an $m$ bit string yields $m$ correct bits with a probability of $0.5 + 0.5\cdot \frac{1}{2^m} \rightarrow 0.5$ for large $m$. Rather than estimating the correct bit values independently by majority votes, we suggest to use the correlations in the data and assign more weight to the data strings which agree with the majority votes in the largest number of bit positions and which are hence most likely to represent correct rather than random bits.

We assume that we have obtained an experiment consisting of $K$ different trials, where $m_k$ qubits have survived in each trial, $k=1\cdots K$. We represent the measurement outcomes of each trial as a string with $n$ entries $(b^k_1,\cdots,b^k_n)$ attaining the values 0 and 1 for the readout of a qubit in state $|0\rangle$ and $|1\rangle$, and for convenience the detection of a missing qubit is represented by the value 0.5 (see the inset in Fig. 3).

For two output strings and $(b^k_1,\cdots,b^k_n)$ and $(b^j_1,\cdots, b^j_n)$, we perform a bit-wise comparison, and introduce the quantifier
\be
\label{chi}
\chi_{kj}(i) =
 \begin{cases}
       1 & (b^k_i, b^j_i) = (0,0)\ \textrm{or}\ (1,1) \\
       -1 & (b^k_i,b^j_i) = (1,0)\ \textrm{or}\ (0,1) \\
       0 & \textrm{otherwise}
     \end{cases}
     \nonumber
\ee

For each trial, we then define a measure for how well all its bit values agree with the values obtained in the all other $(j\neq k$) trials
\be
\label{cind}
C_k = \sum_{j\ne k} \sum_{i=1}^n \chi_{kj}(i).
\ee

The values of $C_k$ fall in two groups: random bits lead to values of $C_k$ fluctuating around $0$, while the positive contributions to $C_k$ for correct bit strings from other correct strings cause a bias of their distribution towards positive values.
Using this, we define a new, weighted estimate for the individual bit-values,
\be
\label{fst}
b_i = \frac{1}{2}\left[1+sgn\left ( \sum_{k} C_k  (b^k_i - \frac{1}{2}) \right )\right].
\ee
The procedure does not suggest a preferred value for bit values when the argument of the $sgn$ function vanishes, but this will occur with smaller and smaller probability the larger the number of trials.

We have simulated trial records with half the qubits remaining on average, and after adequate time to render the density matrix close to the fully mixed state in each relevant two-dimensional $m$ component, such that each string is with almost equal probability a random string and a correct string (in the places of remaining qubits). We have then analyzed numerically the outcome of the filtering algorithm presented, and we show results in Fig. 3.

The calculations are done for a system with $n= 24$ qubits, and the blue curve (circles) shows how many bit values are correctly estimated by a simple majority vote after $K = 10$ trials. This value fluctuates from experiment to experiment, and we show its probability distribution, according to which on average about 83\% of the bits are correct and with about 10\% probability all bit values are correct. The red curve (squares) shows the more advanced filtering, where the trial outcomes are assigned weights as described in Eq. (\ref{fst}). We observe a significant improvement in the average number of correct bits (95\%) and with a 50\% probability all the bit values are now determined correctly. Since the tasks solved by the Grover algorithm often entail a simple method to verify the outcome, the latter scheme presents a very efficient way to obtain the solution of the search problem in the presence of losses.

\begin{figure}
\includegraphics[height=80mm,width=120mm]{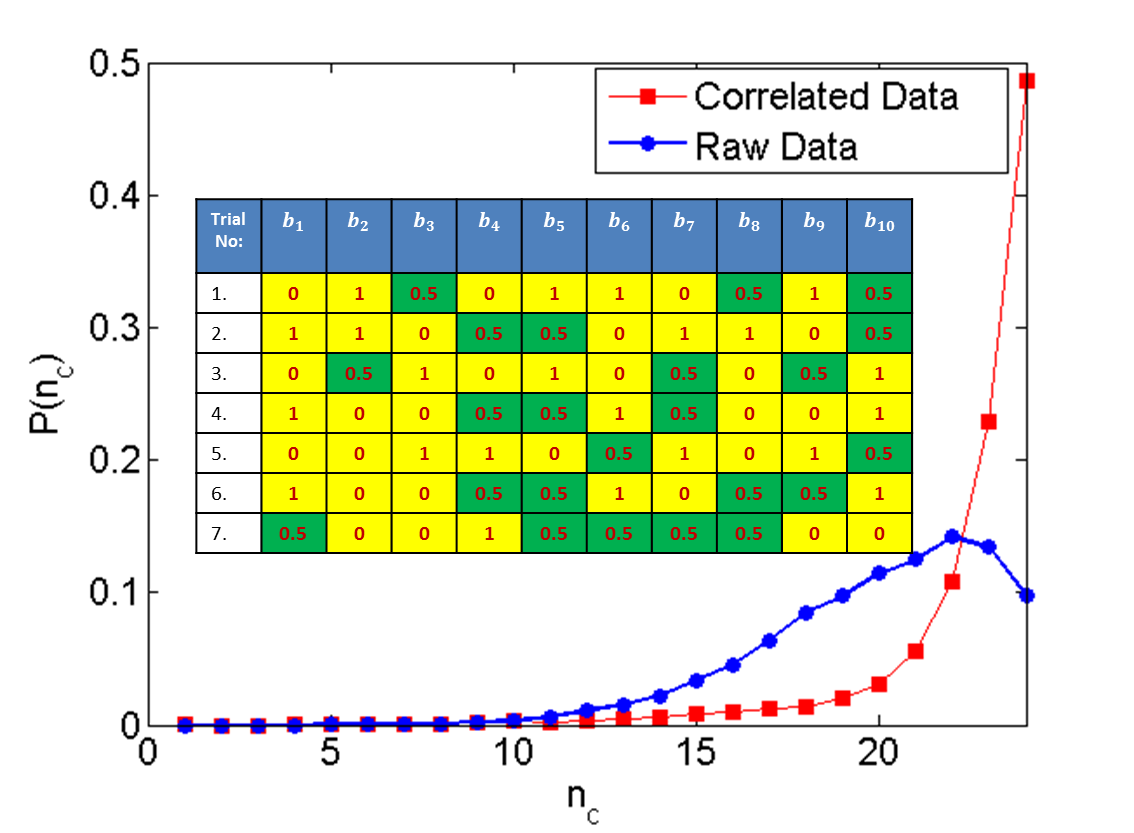}
\caption{The probability distribution of the number $n_c$ of correctly estimated qubits for an initial register size of $n=24$ qubits. In a given experiment we make $10$ trials, to obtain the final bit string, and we compare the cases where the bit values are obtained (i) by a simple majority vote on the raw data (blue circles) (ii) by the correlated data approach (red squares)  using Eq. (\ref{fst}).
To obtain the probability distributions shown, we have averaged the outcome of $10^4$ such experiments.
Inset: The upper left corner of the measurement table displaying the results for each trial. The dark, green boxes indicate the lost qubits, represented by the numerical values $b^k_i=0.5$ in Eqs. (\ref{chi}, \ref{fst}).}
\end{figure}

In the numerical example we lost half of the qubits on average and we included $K=10$ trials in our analysis. With more trials we further reduce the probability that bit values remain unsettled but as indeed, a fraction of the trials predict correct bit values, only a finite number of trials is needed, and the overhead in performing the Grover algorithm that number of times to reconstruct the missing bits, does not change the $\sqrt{N}$ speed-up of the method.

\section{Physical implementation with Rydberg blockade}

It is a key assumptions in the above analysis, that the conditional phase shift operations that make the operator $G$, Eq. (\ref{gop}), work correctly within the subspace of the qubits actually present. This may not be true for a general implementation of quantum computing, where unitary operations on a register are accomplished through sequences of one- and two-bit operations, and where missing physical particles introduce errors that cascade into the time evolution of the ones being still present.

In a recent quantum computing proposal \cite{exp5}, it was shown that the Rydberg blockade interaction \cite{jaksch,laser1,laser2,imp1,imp2} if it extends to any pair of atoms trapped in separate microtraps within a small volume in space, can be used to efficiently implement the Grover algorithm. The basic ideas of the proposal is to use $\pi$-pulses to sequentially excite one atom after the other from a definite ground state or ground state superposition to the Rydberg state. As long as no atom occupies the ground state addressed by the laser pulse no excitation is formed, while the first atom being excited by a pulse will block any further excitation in the sample. Applying sequential excitation $\pi$-pulses followed by sequential de-excitation $\pi$-pulses in the reverse order to the atoms, will return all population to the initial ground state, but a change of sign applies if a single atom was excited. This occurs if any non-zero number of atoms populates a state coupled by the laser field. If all excitation pulses are implemented from the qubit state \emph{different from} the target bit value, this change of sign is readily encoding the Grover conditional phase. I.e., the target state component experiences a change of sign with respect to all other components having one or more erroneous qubits. The sign of the fully symmetric state with respect to all other states is equivalently obtained by $\pi$-pulse excitation and de-excitation sequences from the ``erroneous'' state $(|0\rangle -|1\rangle)/\sqrt{2}$, leaving only the phase of the symmetric state unchanged. For details, see \cite{exp5}.

Missing atoms cannot be excited, and since the above operations work by putting a change of sign on any state, where one or more atoms can be excited from a wrong ground state or ground state superposition, the occasional missing of an atom does not disturb the execution of the algorithm. One may apply the laser pulse sequence as if all atoms are present, and the pulses illuminating empty traps have no effect on the implementation of the desired operations on the remaining atoms.

\section{Conclusion}

In conclusion we have presented here an analysis of Grover search algorithm in the presence of qubit losses.
The effective dynamics in the presence of losses takes place in $n$ two-dimensional subspaces, which allows a simple analytical and numerical investigation of its performance in the presence of losses. We showed that if up to half of the qubits are lost on the time scale of execution of the ideal Grover search, the resulting state of the remaining qubits has a large overlap with the desired target state restricted to the same qubits. We established a correlated data analysis method that performs better than a simple majority vote over repeated experiment, and we demonstrated that the full target state can be reconstructed by repeating the experiment a number of times that scales only linearly with the number of bits in the register. \emph{I.e.}, the $\sqrt{N}$ quantum speed-up is not deteriorated by the losses.

In schemes where it is possible to register when particles are lost, one may adapt any implementation of the Grover algorithm, as long as one adjusts the operations on the system such that the algorithmic steps are always applied correctly to the register composed of the qubits still present.

We presented an experimental implementation scheme with neutral atoms and Rydberg blockade interactions, which naturally lends itself to demonstration of our results, because it can be applied on a lossy physical system without any attention to the losses during the operation of the algorithm. Current experiments with trapped arrays of atoms subject to Rydberg excitation \cite{imp1,imp2, imp3}, may thus implement the protocols discussed in this paper and carry out basic tests of the Grover algorithm on multiple qubits, even beyond the time scale of individual particle losses.

\begin{acknowledgements}
        This work was supported by the EU integrated project AQUTE, the IARPA MQCO program, and DARPA and NSF award PHY-0969883.
        \end{acknowledgements}


\begin{thebibliography}{23}
\bibitem{grov}L.K. Grover, {\it Phys. Rev. Lett.} {\bf 79}, 4709 (1997); {\it Phys. Rev. Lett.} {\bf 80}, 4329 (1998).
\bibitem{gengrov}E. Biham, O. Biham, D. Biron, M. Grassl, and D. A. Lidar, {\it Phys. Rev. A} {\bf 60}, 2742 (1999).
\bibitem{exp1} I. L. Chuang, N. Gershenfeld, and M. Kubinec, {\it Phys. Rev. Lett.}
{\bf 80}, 3408 (1998).
\bibitem{exp2} N. Bhattacharya, H. B. van Linden van den Heuvell, and R. J. C. Spreeuw, {\it Phys. Rev. Lett.} {\bf 88}, 137901 (2002).
\bibitem{exp3}K.-A. Brickman, P. C. Haljan, P. J. Lee, M. Acton, L. Deslauriers, and C. Monroe, {\it Phys. Rev. A} {\bf 72}, 050306(R) (2005) .
\bibitem{exp4}J. Ahn, T. C.Weinacht, and P. H. Bucksbaum, {\it Science} {\bf 287}, 463 (2000).
\bibitem{exp5}K. M{\o}lmer, L. Isenhower, and M. Saffman, {\it J. Phys. B} {\bf 44}, 184016 (2011).
\bibitem{noise1}D. Shapira, S. Mozes, and O. Biham, {\it Phys. Rev. A} {\bf 67}, 042301 (2003).
\bibitem{noise2}N. Shenvi, K. R. Brown, and K. B. Whaley {\it Phys. Rev. A} {\bf 68}, 052313 (2003).
\bibitem{num1}P.H. Song, I. Kim, {\it Eur. Phys. J. D} {\bf 23}, 299 (2003).
\bibitem{num2} R. J. C. Spreeuw and T. W. Hijmans, {\it Phys. Rev. A} {\bf 76}, 022306 (2007).
\bibitem{num3}P. J. Salas, {\it Eur. Phys. J. D} {\bf 46}, 365 (2008).
\bibitem{laser1}M. Saffman and T. G. Walker, {\it Phys. Rev. A} {\bf 72}, 022347 (2005).
\bibitem{laser2}M. Saffman, T. Walker and K. M{\o}lmer {\it Rev. Mod. Phys.} {\bf 82} 2313 (2010).
\bibitem{varnava} M. Varnava, D.E. Browne, and T. Rudolph, {\it Phys. Rev. Lett.} \textbf{ 97}, 120501 (2006).
\bibitem{ralph} T.C. Ralph,  A. J. F.  Hayes and A. Gilchrist, {\it Phys. Rev. Lett.} \textbf{95}, 100501 (2005).
\bibitem{stace} T.M. Stace, S.D. Barrett, and A.C. Doherty, {\it Phys. Rev. Lett.} \textbf{102}, 200501 (2009).
\bibitem{herrera} D. Herrera-Marti and T. Rudolph, arXiv:1208.5148.
\bibitem{jaksch} D. Jaksch, H.-J. Briegel, J. I. Cirac, C. W. Gardiner, and P. Zoller, Phys. Rev. Lett. \textbf{82}, 1975 (1999).
\bibitem{imp1}L. Isenhower, E. Urban, X. L. Zhang, A. T. Gill, T. Henage, T. A. Johnson, T. G. Walker, and M. Saffman, {\it Phys. Rev. Lett.} {\bf 104}, 010503 (2010).
\bibitem{imp2}T. Wilk, A. Ga{\"e}tan, C. Evellin, J. Wolters, Y. Miroshnychenko, P. Grangier, and A. Browaeys, {\it Phys. Rev. Lett.} {\bf 104}, 010502 (2010).
\bibitem{imp3}I. Bloch, J. Dalibard, and W. Zwerger, {\it Rev. Mod. Phys.} {\bf 80}, 885 (2008).
\end{thebibliography}
\end{document}